\documentclass[aps,prl,twocolumn,showpacs,amsmath,amssymb]{revtex4}
\usepackage{graphicx}% Include figure files
\usepackage{dcolumn}% Align table columns on decimal point
\usepackage{bm}% bold math

\begin{document}

\title{Electron-phonon effects and transport in carbon nanotubes}

\author{Vasili Perebeinos, J. Tersoff, and Phaedon Avouris$^*$}
\affiliation{IBM Research Division, T. J. Watson Research Center,
Yorktown Heights, New York 10598}

\date{\today}

\begin{abstract}
We calculate the electron-phonon scattering and binding in semiconducting
carbon nanotubes, within a tight binding model.
The mobility is derived using a multi-band Boltzmann treatment.
At high fields, the dominant scattering is inter-band scattering by
LO phonons corresponding to the corners K of the graphene Brillouin zone.
The drift velocity saturates at approximately
half the graphene Fermi velocity.
The calculated mobility as a function of temperature, electric field, and
nanotube chirality are well reproduced by a simple interpolation formula.
Polaronic binding give a band-gap renormalization of $\sim$70 meV,
an order of magnitude larger than expected.
Coherence lengths can be quite long but are strongly energy dependent.
\end{abstract}

\maketitle

Carbon nanotubes have enabled novel electronic devices,
including quasi-one-dimension field-effect transistors \cite{Avouris}
and electro-optical devices \cite{Misewich,Marcus2}.
Much effort has gone into determining the transport properties,
which are crucial for these and other applications.
Nanotubes exhibit high
mobility at low electric fields, exceeding the best
semiconductors at room temperature \cite{Durkop}. At high fields, the mobility
is dramatically reduced by optical phonon emission, leading to
interesting saturation behavior \cite{Yao,Javey,Park}. However,
little is known about how the mobility varies with temperature,
electric field, and tube diameter and chirality.

Here we calculate the electron-phonon interactions and
drift velocity in semiconducting nanotubes,
within a standard tight-binding approach.
Our results are consistent with available measurements,
and provide a detailed microscopic picture of phonon scattering.
For example, we find that there is a broad maximum in
coherence length at intermediate energies,
centered at 100-150 meV, with coherence lengths of
over a micron at room temperature, and much larger
at low temperature.
At large fields, the mobility is limited primarily
by interband scattering from the LO phonons corresponding
the corners of the graphene Brillouin zone.

At the same time, we provide a broad overview of the transport,
by calculating the mobility over a wide range of temperature,
electric field, and nanotube structure.
The low-field mobility depends strongly on nanotube diameter
and temperature, while at high field the drift velocity approaches
a roughly universal value.
We show that these dependences can be well described by
a simple formula.

In addition, we find that polaronic binding gives a
significant bandgap renormalization, reducing the gap
by around 70 meV over a range of tube diameters.
This is more than an order of magnitude larger than suggested
by previous calculations \cite{Alvez,Piegari},
and is particularly important for larger-diameter tubes,
where it represents a significant fraction of the bandgap.

Our calculations use a standard tight-binding description
for the electronic structure \cite{Saito}.
The phonon energetics are described using an improved model
somewhat similar to that of Aizawa {\it et al.}~\cite{Aizawa}.
We model the electron-phonon (e-ph) interaction by the
Su-Schriefer-Heeger (SSH) model \cite{Su}, with matrix element
$t=t_0-g\delta u$ dependent on the change of the nearest neighbor
C-C distance ($\delta u$), where $t_0=3$ eV. We take the e-ph
coupling constant to be $g=5.3$ eV/\AA~ as predicted theoretically
for a related molecular problem \cite{Perebeinos2}, consistent
with fits to the Peierls gap in conjugated polymers
\cite{Peierls}. The Fourier transformed SSH Hamiltonian
here is
\begin{eqnarray}
{\cal H}_{\rm e-ph}=\sum_{kq\mu}{\rm
M}_{kq}^{\mu}(v^{\dagger}_{k+q}v_{k}-u^{\dagger}_{k+q}u_{k})
(a_{q\mu}+a_{-q\mu}^{\dagger}), \label{eq1}
\end{eqnarray}
where ${\rm M}_{kq}^{\mu}\propto gN^{-1/2}$ is the e-ph
coupling \cite{Mahan}; $u^{\dagger}_{k+q}$ ($v_{k}$) denotes creation
(annihilation) of an electron in the conduction (valence) band;
$a^{\dagger}_{-q\mu}$ is a phonon creation operator with
wavevector $-q$ and phonon band index $\mu=1...6$; $N$ is the
number of the primitive unit cells each containing two carbons.
The indices $k$ and $q$ each label both the continuous
wavevector along the tube axis and the discrete circumferential
wavevector.

Since conduction and valence bands are symmetric in carbon
nanotubes, we consider a filled valence band with a
single electron in the conduction band.
The behavior of a single hole would be identical.
The charge carrier with wavevector $k$ and energy
$\varepsilon_k$ can be scattered to a state with wavevector $k+q$ by
absorbing (emitting) a phonon with wavevector $q$ ($-q$),
and having energy
$\hbar\omega_q$ ($=\hbar\omega_{-q}$) such that the net momentum
and energy are conserved.

Scattering and binding correspond to the imaginary and
real parts of the self-energy.
We calculate these in the Random Phase Approximation.
The scattering rate for an electron of wavevector $k$ is
\begin{eqnarray}
\frac{1}{\tau_k}&=&\sum_{q}W_{k,k+q}
\nonumber \\
W_{k,k+q}&=&\frac{2\pi}{\hbar}\sum_{\mu}\left\vert
M_{k,q}^{\mu}\right\vert^2[n_{q\mu}\delta(\varepsilon_{k}-
\varepsilon_{k+q}+\hbar\omega_{q\mu})
\nonumber \\
&+&(1+n_{-q\mu})\delta(\varepsilon_{k}-
\varepsilon_{k+q}-\hbar\omega_{-q\mu})],
\label{eq2}
\end{eqnarray}
where $\tau_k$ is the scattering time, and the phonon
occupation number $n_q$ is given by the Bose-Einstein distribution.
The energy shift due to electron-phonon coupling is
\begin{eqnarray}
\delta E_k={\rm Re}\sum_{q\mu} && \left\vert
M^{\mu}_{kq}  \right\vert^2
\left[
\frac{n_{-q}+1}{\varepsilon_{k}-\varepsilon_{k+q}-\hbar\omega_{-q\mu}-i\delta}
\right.
\nonumber\\
& +&  \left. \frac{n_{q}}{\varepsilon_{k}-
\varepsilon_{k+q}+\hbar\omega_{q\mu}-i\delta}
\right]  ,
\label{eq3}
\end{eqnarray}
in the limit $\delta\rightarrow 0$.

\begin{figure}
\includegraphics[height=3.63in,width=3.10in]{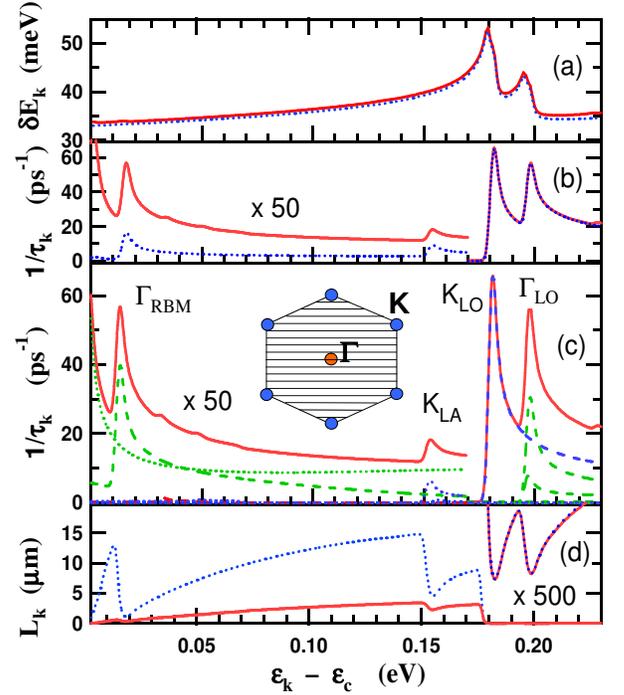}
\caption{\label{figTauE} (color online). Phonon scattering and
binding vs.\  electron energy $\varepsilon_k$ (relative to
conduction band edge $\varepsilon_c$), for a (25,0) tube ($d$=2.0
nm): (a) binding energy, Eq.~(\protect{\ref{eq3}}); (b-c) inverse
lifetime, Eq.~(\protect{\ref{eq2}}); and (d) coherence length. The
energy range shown includes only the first band. In (a), (b), and
(d), solid red curves are for T=300 K, dotted blue curves are for
T=10 K. Note the large change in scale between low and high
energy, $\times$50 in (b) and (c), and $\times$500 in (d). (c)
Decomposition of e-ph scattering rate for 300K: total (red curve);
$\Gamma$-point contributions (green curves); K-point contributions
(blue curves). $\Gamma_{RBM}$ is radial breathing mode. The inset
shows 2D graphene Brillouin zone, with the integration area shown
by the circles of radius equal to 0.1 of the $\Gamma$-M distance,
which contributes virtually 100\% of the total scattering; lines
are only schematic.}
\end{figure}

The results for the binding energy and the scattering rate
of a charge carrier in the first band are shown in
Fig.~\ref{figTauE} for a tubes of diameter  d=2.0 nm.
The binding energy in Fig.~\ref{figTauE}a is nearly independent
of temperature over the entire energy range, and only
weakly energy-dependent at low energy. The binding energy
increases sharply at resonance with the optical phonon $K_{LO}$.
The scattering rate (inverse lifetime) in
Fig.~\ref{figTauE}b shows a strong temperature dependence at low energy,
but is two orders of magnitude larger and nearly independent of temperature
at high energy.

Virtually all of the scattering is by phonons corresponding
to a small region near the $\Gamma$ and $K$ points of the 2D
graphene Brillouin zone, and Fig.~\ref{figTauE}c shows the
contributions of the respective bands to the scattering.
At low electron energy, the scattering is by acoustic phonons
near the $\Gamma$-point.  The lowest-energy phonon band gives
negligible scattering and is not shown.  The
second band is the acoustic phonon contribution which peaks at the
van-Hove singularity (bottom of the conduction band). The third
band is the radial breathing mode which gives a peak labelled
$\Gamma_{\rm RBM}$ at 15 meV in (25,0) tube. The next phonon mode
with a non-negligible e-ph coupling is the longitudinal acoustic
phonon at K-point which gives a weak scattering peak
labelled K$_{\rm LA}$.

By far the strongest coupling is to the LO phonons, which give scattering
roughly two orders of magnitude stronger than the acoustic modes.
The K$_{\rm LO}$ mode corresponds to inter-band scattering.
It is the most important mode at high field, both because
it is strongest, and because it can scatter electrons
before they reach the energy of the $\Gamma_{\rm LO}$ mode.

In Fig.~\ref{figTauE}d we show the coherence length, defined as
$L_k=v_k\tau_k$, where $v_k=d\varepsilon_k/d k$ is the band
velocity. (We approximate this by the bare velocity
$v_k=\partial\varepsilon_k/\partial k$, which we calculate to be
the same to within a few percent except near resonance with
the optical phonons K$_{\rm LO}$ and $\Gamma_{\rm LO}$.)
Note that the coherence length is strongly energy-dependent,
with a broad maximum between the breathing mode and the
optical phonons.  It is also quite sensitive to temperature,
but even at room temperature the coherence length
can be well over a micron, consistent with lengths inferred from
experiment \cite{Javey,Park}.  Charge carriers can be
injected into nanotubes well above the band edge \cite{Heinze03},
suggesting the possibility of ballistic or even quantum-coherent
devices over a range of length scales.
At higher energy, the carriers
have a temperature independent coherence length of around 20-40 nm,
depending on energy. This is consistent with the lengths
inferred from experiments for metallic tubes \cite{Yao,Javey,Park}.

For applications such as device modeling,
what is needed is the total effect of scattering on the transport.
This is expressed as a mobility, which we calculate by
solving the steady-state multi-band Boltzmann equation
in the presence of an electric field.
The electron momentum distribution function $g_k$ is
\begin{eqnarray}
\frac{eE}{\hbar }\frac{\partial g_{k}}{\partial k}=-\sum_{q}&&[
W_{k,k+q}g_{k}\left( 1-g_{k+q}\right)
\nonumber\\
&&-W_{k+q,k}g_{k+q}\left( 1-g_{k}\right)], \label{eq4}
\end{eqnarray}
where $W_{k,k+q}$ is given in Eq.~(\ref{eq2}). Using the
non-equilibrium distribution function $g_k$ from Eq.~(\ref{eq4}),
we calculate the drift velocity $v$ and the mobility $\mu$:
\begin{eqnarray}
v=\sum_kg_k\frac{\partial \varepsilon_{k}}{\hbar\partial k}=\mu E
\label{eq5}
\end{eqnarray}

\begin{figure}
\includegraphics[height=2.96in,width=3.34in]{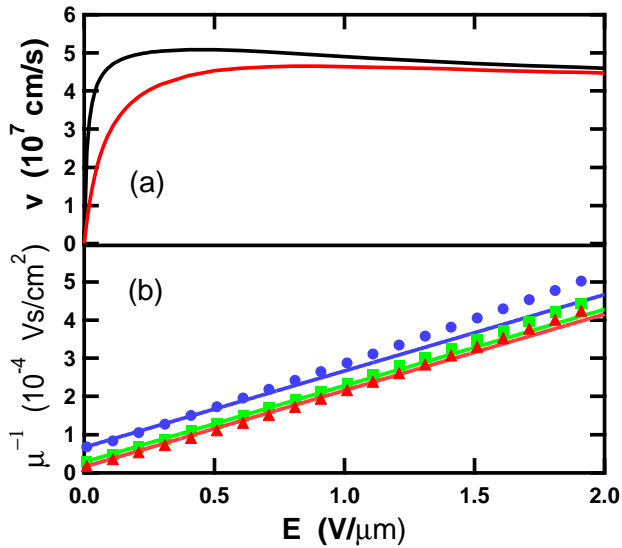}
\caption{\label{figVMuField} (color online). (a) Drift velocity
vs. electric field at T=300 K (red curve) and T=10 K (black curve)
in (25,0) tube. (b) Inverse mobility at T=300 K vs. $E$-field;
straight lines show corresponding fits to
Eq.~(\protect{\ref{eq6}}), with $v_s$=5.0$\times$10$^7$. Fitted
values of $\mu_0$ are $\mu_0$=65000 cm$^2$/Vs for (25,0) tube (red
triangles); $\mu_0$=35500 cm$^2$/Vs for (19,0) tube (green
squares); and $\mu_0$=15000 cm$^2$/Vs for (13,0) tube (blue
circles).}
\end{figure}

The drift velocity versus $E$-field is shown in Fig.~\ref{figVMuField}a,
for a (25,0) tube. We show results at room temperature and at
low temperatures, in the limit of low carrier density.
The drift velocity saturates for fields $E \gtrsim 0.5$ V/$\mu$m.
At low temperature we find negative differential resistance,
i.e.\  drift velocity decreasing with increasing field.
A similar observation was reported previously \cite{Pennington},
and it was speculated to result from population of the second band
at high field.
However, we find that the behavior persists even if we restrict
the basis set to include only the first band.

The inverse mobility shown in Fig.~\ref{figVMuField}b can be fitted
rather well by a simple linear function of the $E$-field,
over a wide range of  applied E-fields:
\begin{eqnarray}
\mu^{-1} \approx \mu_0^{-1}+v_s^{-1}E, \label{eq6}
\end{eqnarray}
where $\mu_0$ is the zero-field mobility and $v_s$ is the
saturation velocity. This is analogous to the linear dependence
$R=R_0+V/I_0$ found for the dependence of the resistance $R$  of
metallic nanotubes with applied voltage $V$
\cite{Yao}, where $I_0$ is a saturation current.
The saturation drift velocity
$v_s\approx5.0\pm 0.3\times 10^7$ cm/s is nearly independent
of the tube diameter and temperature.
The lack of strong dependence on diameter is perhaps surprising,
but this velocity is intriguingly close to half of the
maximum band velocity (the Fermi velocity of graphene,
$9.8 \times 10^7$ cm/s for the value of the tight-binding
matrix element $t$ used here).

\begin{figure}
\includegraphics[height=3.20in,width=3.10in]{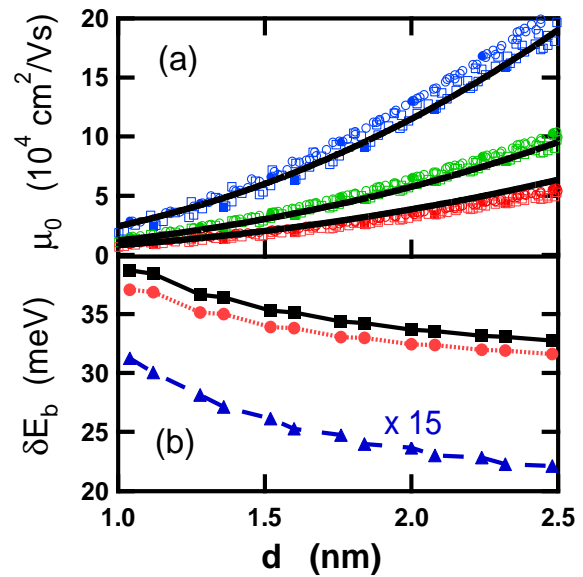}
\caption{\label{figMu0} (color online). (a) Zero-field mobility
vs.\ tube diameter, for tubes of many different chiralities.
Temperatures are  T=450 K (red), T=300 K (green), and T=150 K
(blue). Values are based on linear extrapolation of $\mu^{-1}$,
Eq.~(\ref{eq6}),  for $E$-field between 0.01 and 0.5 V/$\mu$m.
Black curves are fitted to Eq.~(\ref{eq7}). (b) Polaron binding
energy (black squares) vs.\  tube diameter. Red circles show
portion of binding energy from optical phonons, and blue triangles
show the much smaller binding energy ($\times$15) from acoustic
phonons.}
\end{figure}

One of the greatest obstacles to fabricating reproducible nanotube
devices, is the difficulty of controlling (or even knowing) the
tube structure.  This structure, the diameter and chirality, is
specified by two indices (m,n). In Fig.~\ref{figMu0}a, we show how
the zero-field mobility depends on tube structure.
We find that the dependence on diameter and temperature can be
described by a single simple empirical relation:
\begin{eqnarray}
\mu_0 (T,d)=\mu_1\frac{300 {\rm K}}{T}\left(\frac{d}{1 {\rm
nm}}\right)^{\alpha}, \label{eq7}
\end{eqnarray}
with $\mu_1=12000$ cm$^2$/Vs and $\alpha=2.26$.
(The arbitrary constants 300K and 1 nm are simply to give
$\mu_1$ units of mobility, independent of $\alpha$.) The structure
dependence is not entirely captured by the diameter, nor is the
1/T temperature dependence very precise, but overall the agreement
is quite good. Thus the dependence of mobility on temperature,
field strength, and even nanotube structure can be captured to
useful accuracy ($\sim$10\%) with a simple 3-parameter
description, Eqs.~(\ref{eq6}-\ref{eq7}).

\begin{figure}
\includegraphics[height=2.96in,width=3.34in]{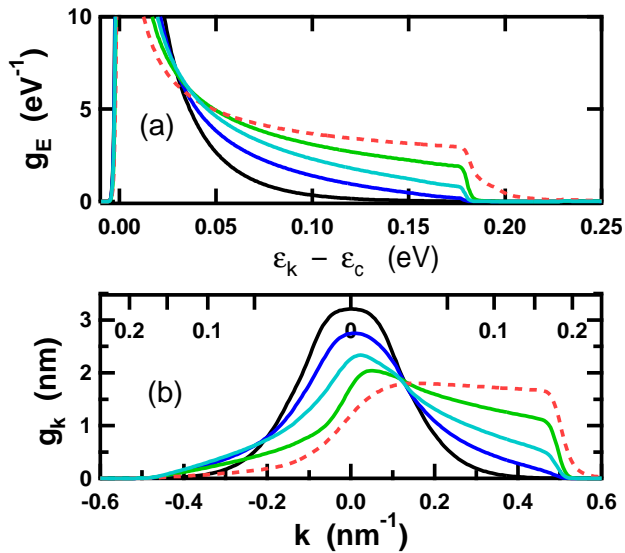}
\caption{\label{figgdist} (color online). Steady-state
distribution function for an electron in a (25,0) tube at T=300 K.
E-fields (from bottom to top on right side) are E=0.001 (black),
0.01 (blue), 0.03 (light blue), 0.1 (green), and 0.5 (red broken
curve) V/$\mu$m. Distributions are shown with respect to (a)
energy, and (b) wavevector along tube axis. (Curves include
Gaussian broadening of 2 meV.) The top axis in (b) shows
corresponding energy.}
\end{figure}

It might seem surprising that, while the scattering
in Fig.~\ref{figTauE} is strongly energy-dependent,
the mobility in Fig.~\ref{figVMuField} varies smoothly
with increasing field.
This can be understood by considering the distribution function
$g_k$ [or equivalently $g_E$] obtained by solving
the Boltzmann equation.  This is shown in Fig.~\ref{figgdist}.
At very low field, the scattering is entirely by acoustic phonons,
giving a smooth distribution in $k$. (The distribution in energy
reflects the van Hove singularity in the density of states.) With
increasing field, the electron has an increasing probability of
reaching the optical phonon energy, at which point it is quickly
scattered. This leads to a step in the distribution near the
phonon energy (0.18 eV). With increasing field, the distribution
becomes flatter below this energy, until at high field the
distribution is almost flat in $k$, with a sharp step at the
phonon energy. From Eq.~(\ref{eq6}), we might expect the crossover
from acoustic to optical phonons as the dominant scattering
mechanism to occur roughly at fields $E \sim v_s / \mu_0$,
corresponding to $\sim$0.08 V/$\mu$m in Fig.~\ref{figgdist}. This
is indeed roughly the field strength where intermediate behavior
is seen in $g_k$. Then from Eqs.~(\ref{eq6}-\ref{eq7}), we can
anticipate that the crossover field varies with temperature and
tube size approximately as $d^{-\alpha} T$.

Finally we return to the polaronic energy shift.
Figure~\ref{figMu0}b shows the shift calculated with
Eq.~(\ref{eq3}), as a function of nanotube diameter. (This refers
to the band-edge states, $\varepsilon_k$=$\varepsilon_c$ in
Fig.~\ref{figTauE}a.) We find binding energies of around 35 meV,
with only a weak dependence on diameter over the typical range.
This corresponds to a band-gap renormalization of 70 meV due to
electron-phonon interactions, which is quite significant for
large-diameter tubes. This is more than an order of
magnitude larger than previous theoretical estimates
\cite{Alvez,Piegari}. The reason for this discrepancy is explained
in Fig.~\ref{figMu0}b. Previous calculations used a continuum
model, which is equivalent to including acoustic but not optical
phonons. However, almost all the binding is due to optical
phonons, which couple far more effectively to the electrons. If we
repeat the calculations using only the first three phonon bands, we
obtain only a weak binding, in excellent agreement with continuum
calculations \cite{Piegari}.

In conclusion, we find that in semiconducting carbon nanotubes,
the scattering of electrons or holes by phonons
is strongly energy dependent, both at low energy
and around the optical phonon energies.
Nevertheless, the mobility at low carrier densities
can be described fairly accurately by a simple formula,
which should be useful in the analysis of nanotube devices.
The polaronic binding energy is much larger than expected,
giving a significant renormalization of the bandgap.

\newpage

\end{document}